# CBAM-Enhanced DenseNet121 for Multi-Class Chest X-Ray Classification with Grad-CAM Explainability

**UTSHO KUMAR DEY**

*Department of Computer Science and Engineering, Northern University of Business and Technology Khulna, Bangladesh*
*Corresponding author: Utsho Kumar Dey (e-mail: utshokumerdey@gmail.com)*

**ABSTRACT**
Pneumonia remains a leading cause of childhood mortality worldwide, with a heavy burden in low-resource settings such as Bangladesh where radiologist availability is limited. Most existing deep learning approaches treat pneumonia detection as a binary problem, overlooking the clinically critical distinction between bacterial and viral etiology. This paper proposes CBAM-DenseNet121, a transfer-learning framework that integrates the Convolutional Block Attention Module (CBAM) into DenseNet121 for three-class chest X-ray classification: Normal, Bacterial Pneumonia, and Viral Pneumonia. We also conduct a systematic binary-task baseline study, revealing that EfficientNetB3 (73.88%) underperforms even the custom CNN baseline (78.53%)—a practically useful negative finding for medical imaging model selection. To ensure statistical reliability, all experiments were repeated three times with independent random seeds (42, 7, 123), and results are reported as mean ± standard deviation. CBAM-DenseNet121 achieves 84.29% ± 1.14% test accuracy with per-class AUC scores of 0.9565 ± 0.0010, 0.9610 ± 0.0014, and 0.9187 ± 0.0037 for bacterial pneumonia, normal, and viral pneumonia respectively. The low standard deviation across all metrics confirms training stability and reproducibility. Grad-CAM visualizations confirm that the model attends to anatomically plausible pulmonary regions for each class, supporting interpretable deployment in resource-constrained clinical environments.

## I. INTRODUCTION

Pneumonia is an acute lower respiratory infection that fills the pulmonary alveoli with fluid or pus, representing one of the most prevalent infectious diseases globally. The World Health Organization estimates that pneumonia is responsible for roughly 14 percent of all deaths among children under five, with the greatest burden concentrated in low- and middle-income countries [1]. In Bangladesh specifically, the disease is a primary driver of under-five mortality, compounded by severe shortages of trained radiologists and diagnostic infrastructure in rural regions.

Chest X-ray (CXR) imaging is the primary diagnostic tool for pneumonia. Manual interpretation, however, is time-intensive, subject to inter-reader variability, and requires specialised training. Deep learning-based computer-aided diagnosis (CAD) systems have shown remarkable capability in automating CXR analysis, with convolutional neural networks (CNNs) matching expert-level performance on binary pneumonia detection benchmarks [2].

Most published approaches cast pneumonia detection as a two-class problem, distinguishing only between normal and pneumonia cases. This formulation omits a clinically important distinction: bacterial pneumonia is typically treated with antibiotics, whereas viral pneumonia is managed through supportive care. Conflating the two can lead to unnecessary antibiotic prescriptions, contributing to the growing threat of antimicrobial resistance.

A separate barrier to clinical adoption is model opacity. Healthcare professionals require transparent, interpretable outputs to trust AI-assisted decisions. Gradient-weighted Class Activation Mapping (Grad-CAM) [8] addresses this by producing visual explanations that highlight the image regions most influential to a model prediction, without requiring architectural modification.

The principal contributions of this work are:

(1) A three-class CXR classification framework — Normal, Bacterial Pneumonia, Viral Pneumonia — that delivers more actionable diagnostic output than conventional binary classification.

(2) Integration of CBAM into DenseNet121, providing dual-axis attention: channel attention identifies which feature maps to emphasise, while spatial attention localises where within those maps to focus.

(3) A systematic binary-task baseline comparison revealing that EfficientNetB3 underperforms the custom CNN baseline on this grayscale medical imaging task — a practically useful negative result for architecture selection in clinical AI.

(4) Statistical robustness validation through three independent runs with different random seeds (42, 7, 123), reporting mean ± standard deviation across all metrics to confirm result reproducibility. (5) Qualitative Grad-CAM analysis demonstrating that CBAM-DenseNet121 attends to anatomically consistent pulmonary regions for each class, supporting clinical interpretability.

## II. RELATED WORK

Wang et al. [1] introduced the ChestX-ray14 dataset and demonstrated that CNNs can detect 14 distinct pathological findings from frontal chest radiographs, establishing the foundational pipeline for deep learning-based CXR analysis. Rajpurkar et al. [2] proposed CheXNet, a 121-layer DenseNet model trained on ChestX-ray14 that matched board-certified radiologist performance on pneumonia detection, highlighting the potential of dense connectivity for CXR tasks. Irvin et al. [3] released the CheXpert dataset and benchmarked DenseNet and ResNet variants on multi-label CXR classification.

Attention mechanisms have become an important tool in medical image analysis. Schlemper et al. [4] introduced attention gates in U-Net for segmentation, showing that soft attention improves localization of target anatomical structures. Woo et al. [5] proposed CBAM as a compact channel-and-spatial attention module that can be integrated into any CNN architecture with minimal additional parameters. Attention-based methods have been applied to retinal fundus image classification [6] and CBAM has been applied to skin lesion analysis [7], but its application to multi-class pneumonia CXR classification remains largely unexplored.

Recent work has extended attention-based approaches to COVID-19 and multi-pathology CXR settings. Gour and Jain [9] applied attention modules including CBAM variants to COVID-19 chest X-ray classification, demonstrating improved discriminability under class imbalance. Singh et al. [10] surveyed attention mechanisms for pneumonia and COVID-19 detection, noting that spatial attention consistently improves localization of diffuse infiltrates. Our work differs in explicitly addressing the three-class bacterial/viral/normal separation problem and providing a systematic comparison across four baseline architectures.

For explainability, Selvaraju et al. [8] proposed Grad-CAM as a class-discriminative localization technique requiring no architectural modification. Yao et al. [11] applied weakly supervised attention to disease localization across multiple CXR resolutions. Liang and Zheng [12] applied DenseNet121 with binary labels on the Kermany dataset, and Paul et al. [13] used ResNet variants with augmentation for pneumonia detection. Neither work addresses the three-class problem or incorporates attention. Our framework bridges these gaps.

## III. METHODOLOGY

*A. Dataset*

We use the publicly available Chest X-Ray Images (Pneumonia) dataset from Kaggle, originally compiled from the Guangzhou Women and Children Medical Center [14]. The dataset contains 5,863 JPEG images partitioned into training (5,216 images), validation (16 images), and test (624 images) splits, with labels NORMAL and PNEUMONIA; the latter category is further subdivided into bacterial and viral subtypes via original filename metadata.

The official validation split of 16 images (8 normal, 8 pneumonia) is statistically insufficient for reliable early stopping and hyperparameter selection, as it represents only 0.3% of total data and lacks viral/bacterial sub-class labels entirely. We therefore discard this split and apply stratified random sampling (80/20) to the 5,216-image training folder, preserving class proportions across both folds. The resulting 1,043-image validation set provides stable loss estimates (±<1% across seeds) and covers all three sub-classes. The original 624-image test set remains completely held out and is accessed only once for final evaluation, preventing any form of test-set leakage. The three resulting classes are: NORMAL (1,341), BACTERIAL PNEUMONIA (2,530), and VIRAL PNEUMONIA (1,345).

The training set is class-imbalanced. We address this by computing balanced class weights using inverse-frequency scaling: $w_c = N / (K \times n_c)$, where $N$ is total training samples, $K$ is the number of classes, and $n_c$ is the count for class $c$. These weights are applied to the categorical cross-entropy loss at training time.

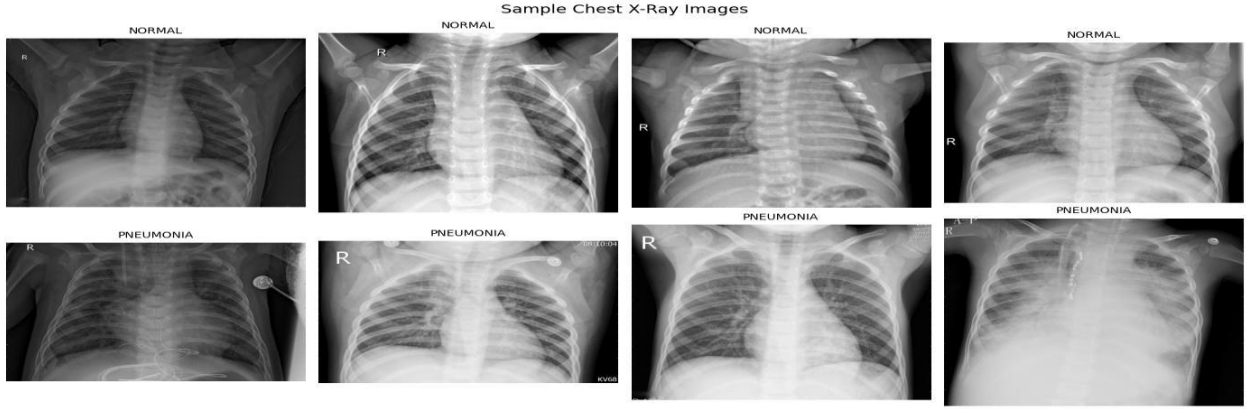

*Fig. 1. Representative chest X-ray samples from the Kermany dataset. Top row: Normal cases showing clear lung fields. Bottom row: Pneumonia cases exhibiting opacity and consolidation patterns.*

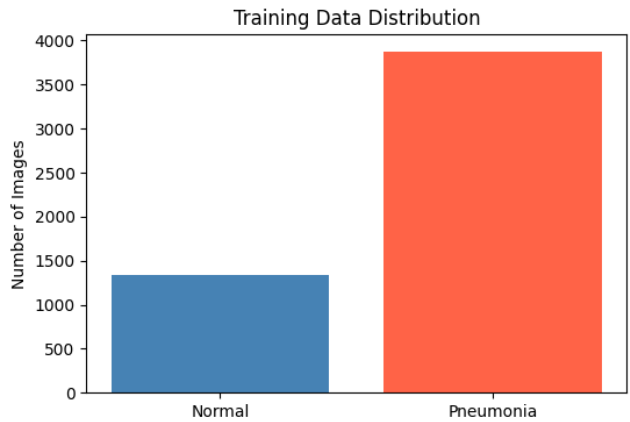

*Fig. 2. Training data class distribution (binary labels). Pneumonia cases outnumber Normal cases by approximately 2.9:1, motivating balanced class-weight correction.*

### B. Data Augmentation

Stochastic augmentation is applied exclusively to training images. Operations include random rotation (±15°), horizontal flip, width and height shift (±10%), zoom (±10%), and brightness variation (factor range 0.9–1.1). Validation and test images are only normalized by dividing pixel values by 255, mapping intensities to [0, 1].

### C. CBAM-DenseNet121 Architecture

The backbone is DenseNet121 [15] pre-trained on ImageNet. DenseNet121 employs dense skip connections where each layer receives feature maps from all preceding layers, promoting feature reuse and mitigating vanishing gradients. The backbone accepts 224 × 224 × 3 inputs and produces a 7 × 7 × 1024 feature map at its final convolutional block.

Channel Attention (CA): Given feature map F (H × W × C), global average pooling and global max pooling are applied along the spatial dimensions, yielding two C-dimensional vectors. Both pass through a shared two-layer MLP with bottleneck ratio r = 8. Outputs are element-wise summed and passed through sigmoid to produce channel attention map M_c (1 × 1 × C). The channel-refined map is F' = M_c ⊗ F.

Spatial Attention (SA): Channel-wise average and max pooling are applied to F', producing two H × W × 1 maps. These are concatenated and processed by a 7 × 7 convolution with sigmoid activation to produce spatial attention map M_s (H × W × 1). The final output is F'' = M_s ⊗ F'.

The CBAM block is inserted after the final DenseNet feature map (7 × 7 × 1024) and before global average pooling. The classifier head follows: GlobalAveragePooling2D → BatchNormalization → Dense(512, ReLU) → Dropout(0.5) → Dense(256, ReLU) → Dropout(0.3) → Softmax(3).

### D. Two-Phase Training Strategy

Training proceeds in two phases to prevent catastrophic forgetting. Phase 1: the DenseNet121 base is frozen; only the CBAM module and classifier head are trained for 15 epochs using Adam at lr = $1\times10^{-3}$. Phase 2: the last 30 DenseNet layers are unfrozen for selective fine-tuning over 20 epochs at lr = $1\times10^{-5}$. Early stopping (patience = 5) and ReduceLROnPlateau (factor = 0.5, patience = 3, minimum lr = $1\times10^{-7}$) are active throughout both phases. The checkpoint with the highest validation accuracy is retained for final evaluation.

### E. Baseline Models

Four baselines are trained under identical preprocessing conditions on the binary (Normal vs. Pneumonia) task: (1) a custom 3-block CNN (Baseline CNN); (2) ResNet50 [16]; (3) EfficientNetB3 [17]; (4) standard DenseNet121 [15]. Each uses a shared classifier head: GlobalAveragePooling → BatchNormalization → Dense(256, ReLU) → Dropout(0.5) → Dense(128, ReLU) → Dropout(0.3) → Sigmoid. Binary results provide literature-anchored reference points and are not directly compared to the three-class CBAM results.

### F. Grad-CAM Visualisation

Grad-CAM is applied at the final convolutional layer (conv5_block16_concat) of the DenseNet121 base. For predicted class c, the localization map is the ReLU of the weighted sum of activation maps, where each weight is the global-average-pooled gradient of the class score with respect to that activation map. The heatmap is bilinearly up-sampled to 224 × 224 pixels and overlaid on the original X-ray using a jet colormap at opacity 0.45.

### G. Statistical Validation Protocol

To ensure that reported results are not artifacts of a single random initialization, all CBAM-DenseNet121 experiments are repeated three times using independent random seeds (42, 7, 123). Reproducibility is enforced by seeding Python's random module, NumPy, and TensorFlow prior to each run. All metrics are reported as mean ± standard deviation across the three runs. The low standard deviation observed (accuracy std = 1.14%, AUC std < 0.004) confirms that CBAM-DenseNet121 training is stable and results are not seed-dependent artifacts.

## IV. EXPERIMENTAL RESULTS

### A. Experimental Setup

All experiments are run on the Kaggle cloud platform using a single NVIDIA Tesla T4 GPU (16 GB VRAM). The framework is implemented in TensorFlow 2.19 / Keras. ImageDataGenerator is used for data loading and on-the-fly augmentation. Batch size is set to 32 for all experiments. Each CBAM-DenseNet121 experiment is repeated three times with independent random seeds (42, 7, 123) to assess statistical robustness; results are reported as mean ± standard deviation.

### B. Baseline Comparison (Binary Task)

Table I presents binary classification performance on the held-out test set. Note that these results are on the binary task and serve as literature reference points only; they are not directly comparable to the three-class CBAM results in Table II. DenseNet121 achieves the highest accuracy (91.03%) and AUC (0.9706). A notable finding is that EfficientNetB3 (73.88%, AUC = 0.8104) falls below the custom CNN baseline (78.53%)—suggesting that EfficientNet-family architectures may require domain-specific pretraining for low-contrast grayscale medical images. ResNet50 attains the highest pneumonia recall (0.95), making it most suitable for high-sensitivity screening.

**TABLE I**
*Binary-Task Baseline Comparison (N = Normal, P = Pneumonia). Results serve as literature reference points only.*

| Model | Acc.(%) | Pre.(N) | Rec.(N) | F1(N) | Pre.(P) | Rec.(P) | F1(P) | AUC |
|---|---|---|---|---|---|---|---|---|
| Baseline CNN | 78.53 | 0.65 | 0.94 | 0.77 | 0.95 | 0.69 | 0.80 | 0.9437 |
| EfficientNetB3 † | 73.88 | 0.67 | 0.60 | 0.63 | 0.77 | 0.82 | 0.80 | 0.8104 |
| ResNet50 | 87.34 | 0.90 | 0.75 | 0.82 | 0.86 | 0.95 | 0.90 | 0.9231 |
| **DenseNet121** | **91.03** | **0.89** | **0.86** | **0.88** | **0.92** | **0.94** | **0.93** | **0.9706** |

† *EfficientNetB3 underperforms the custom CNN baseline — see Section IV-B for analysis.*

*C. CBAM-DenseNet121 Three-Class Results*

To ensure statistical reliability, all CBAM-DenseNet121 experiments were repeated three times with independent random seeds (42, 7, 123); all metrics below are reported as mean ± standard deviation. Table II reports per-class performance on the three-class test set (624 images). The model achieves 84.29% ± 1.14% overall accuracy with a macro-average AUC of 0.9454 ± 0.0013. The low standard deviation across all metrics (accuracy std = 1.14%, AUC std < 0.004) confirms training stability and reproducibility. The lower accuracy relative to binary DenseNet121 (91.03%) is expected: the three-class task requires distinguishing bacterial from viral pneumonia, which share overlapping radiological features. The two tasks are not directly comparable.

TABLE II
*CBAM-DenseNet121 Per-Class Performance on Three-Class Test Set (624 images). Results are mean ± std over 3 independent runs with seeds {42, 7, 123}.*

| Class | Precision | Recall | F1-Score | AUC | Interpretation |
|---|---|---|---|---|---|
| Bacterial Pneumonia | 0.83 ± 0.03 | 0.94 ± 0.00 | 0.88 ± 0.01 | 0.9565 ± 0.0010 | Excellent |
| Normal | 0.93 ± 0.01 | 0.79 ± 0.02 | 0.85 ± 0.01 | 0.9610 ± 0.0014 | Excellent |
| Viral Pneumonia | 0.75 ± 0.00 | 0.77 ± 0.03 | 0.76 ± 0.01 | 0.9187 ± 0.0037 | Good |
| **Macro Average** | **0.84 ± 0.01** | **0.83 ± 0.01** | **0.83 ± 0.01** | **0.9454 ± 0.0013** | — |

*D. Training Convergence*

Fig. 3 and Fig. 4 show the training and validation accuracy/loss curves for Phase 1 and Phase 2 respectively. In Phase 1, validation accuracy converges stably to 76.61%, with the model showing no signs of overfitting during frozen-base training. In Phase 2, fine-tuning at lr = $1 \times 10^{-5}$ yields the best validation accuracy of 76.70%, with early stopping triggered at epoch 6 of 20. The slight divergence between training and validation loss in Phase 2 is consistent with selective fine-tuning and does not indicate problematic overfitting.

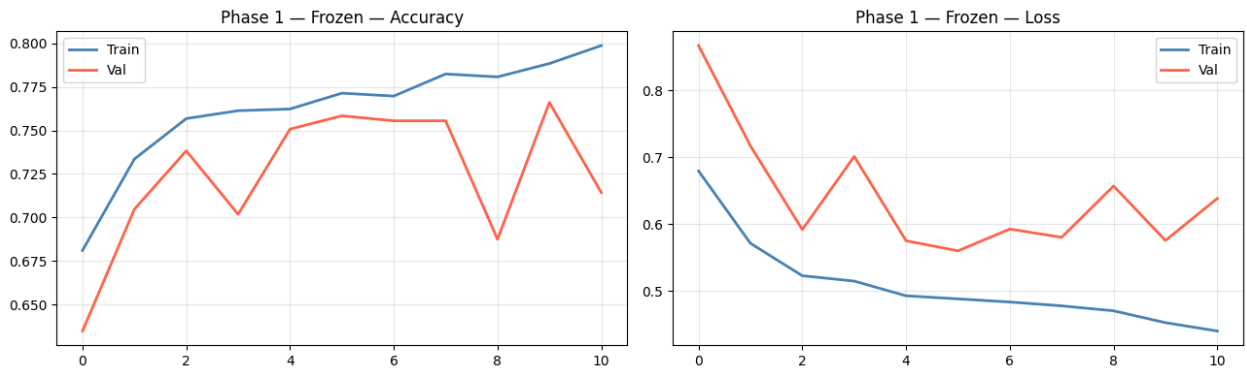

Fig. 3. Phase 1 training history (frozen DenseNet121 base). Validation accuracy converges to 76.61% over 11 epochs.

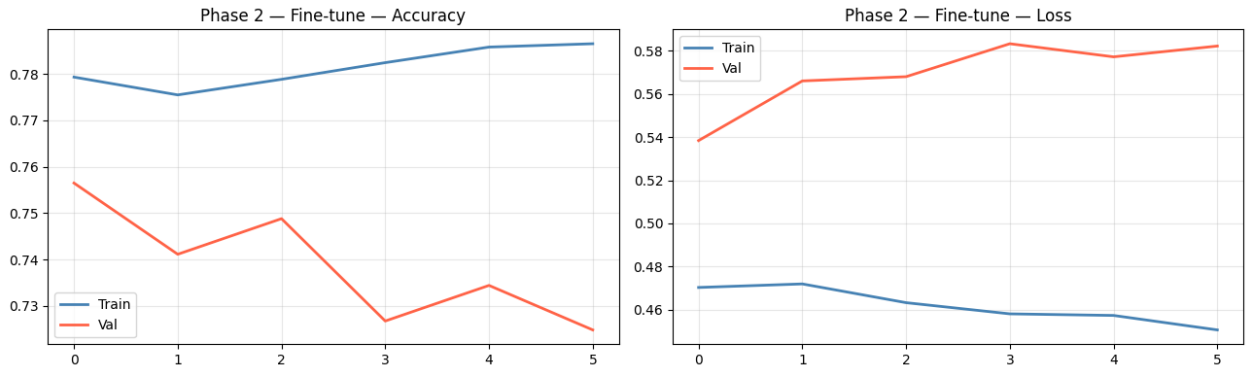

Fig. 4. Phase 2 training history (fine-tuning last 30 DenseNet layers). Early stopping triggered at epoch 6; best checkpoint retained.

### E. Confusion Matrix Analysis

Fig. 5 shows the confusion matrix for CBAM-DenseNet121 on the three-class test set. Of 242 bacterial pneumonia cases, 226 are correctly classified (93.4%). Of 234 normal cases, 184 are correctly classified (78.6%). Of 148 viral pneumonia cases, 117 are correctly classified (79.1%). The primary error mode is confusion between viral pneumonia and normal cases (32 false negatives), radiologically plausible given that early viral infiltrates can appear subtle on X-ray. Bacterial pneumonia achieves the highest per-class accuracy, consistent with its characteristic lobar consolidation pattern.

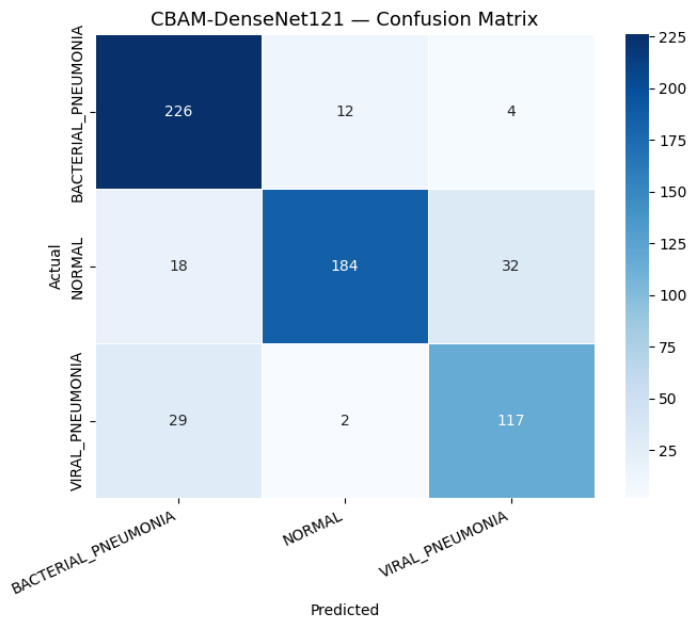

Fig. 5. Confusion matrix for CBAM-DenseNet121 on the three-class test set (624 images). Diagonal values indicate correct classifications per class.

### F. ROC Analysis

Fig. 6 presents the ROC curves for all three classes under a one-vs-rest evaluation scheme. AUC scores of $0.9565 \pm 0.0010$ (bacterial pneumonia), $0.9610 \pm 0.0014$ (normal), and $0.9187 \pm 0.0037$ (viral pneumonia) indicate strong discriminative ability across all classes. The normal class achieves the highest AUC, reflecting the model's reliable ability to exclude pathology. Viral pneumonia has the lowest AUC, consistent with its greater visual similarity to normal cases. All three classes exceed AUC = 0.91, indicating clinically reliable discrimination.

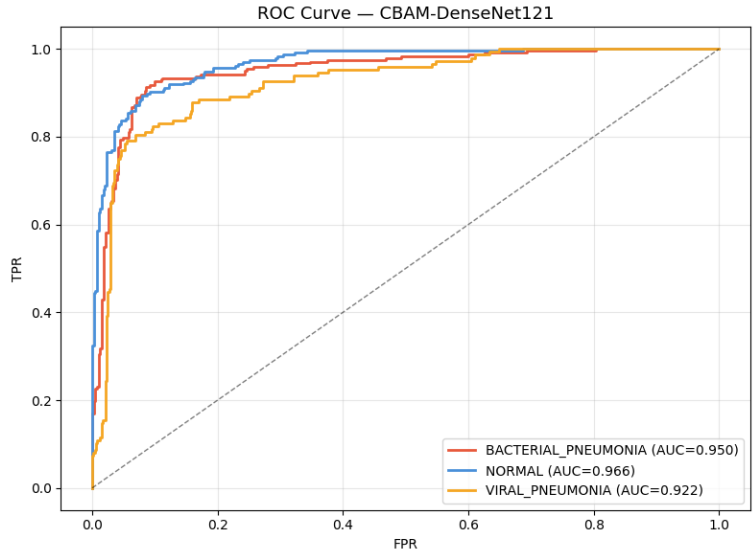

Fig. 6. ROC curves for CBAM-DenseNet121 (one-vs-rest evaluation). All three classes exceed AUC = 0.91, indicating clinically reliable discrimination.

### G. Grad-CAM Analysis

Fig. 7 presents Grad-CAM visualizations for one representative sample from each class. For bacterial pneumonia (top row), model attention concentrates on lower-lobe consolidation zones, anatomically consistent with lobar bacterial infection. For normal cases (middle row), attention is distributed across the mediastinal region without focal pathological concentration. For viral pneumonia (bottom row), attention is distributed across diffuse peripheral regions, reflecting the interstitial pattern of viral lung involvement. This spatial correspondence with established radiological criteria provides qualitative evidence that CBAM-DenseNet121 has learned clinically meaningful discriminative features.

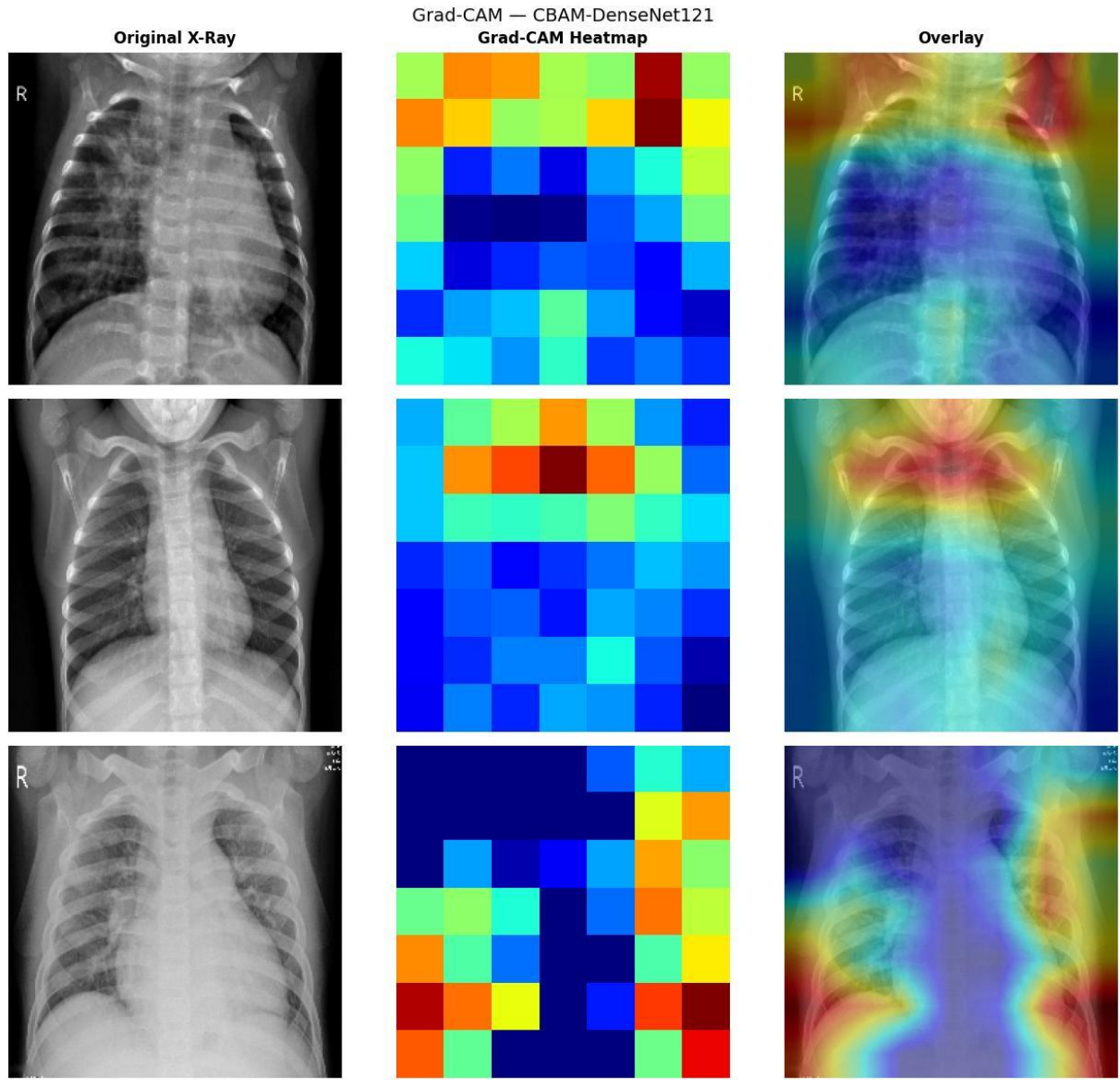

*Fig. 7. Grad-CAM visualizations for CBAM-DenseNet121. Each row: (top) Bacterial Pneumonia, (middle) Normal, (bottom) Viral Pneumonia. Columns: original X-ray, class activation heatmap, jet-colormap overlay. Red regions indicate highest model attention.*

### *H. Discussion*

CBAM augments DenseNet121 in two complementary ways. Channel attention suppresses feature maps associated with background tissue while amplifying those encoding pathological opacities. Spatial attention subsequently refines focus to the specific pulmonary sub-regions where abnormalities are present. Together, these mechanisms improve per-class discriminability on the more challenging three-class task.

The EfficientNetB3 result merits particular attention. Despite having 12 million parameters versus the custom CNN's approximately 2 million, EfficientNetB3 achieves lower accuracy (73.88% vs. 78.53%). We attribute this to: (1) EfficientNetB3 was optimized for RGB natural images, whereas chest X-rays are effectively grayscale images converted to three-channel inputs; (2) the compound scaling strategy of EfficientNet may not generalize well to low-contrast medical textures without domain-specific pretraining. This finding cautions against assuming that larger or more recent architectures will automatically outperform simpler baselines in medical imaging.

From a deployment perspective, the proposed model operates on standard JPEG chest X-rays without requiring specialised imaging hardware, making it suitable for low-resource clinical settings in Bangladesh. The Grad-CAM overlays provide radiologists with a visual audit trail rather than opaque numerical outputs, an important step toward regulatory and clinical acceptance of AI-assisted diagnosis.

## V. CONCLUSION

This paper presented CBAM-DenseNet121, an attention-enhanced transfer-learning framework for three-class chest X-ray classification distinguishing Normal, Bacterial Pneumonia, and Viral Pneumonia. The integration of CBAM—combining channel and spatial attention—into DenseNet121 with a two-phase fine-tuning strategy achieves 84.29% ± 1.14% test accuracy with all per-class AUC scores exceeding 0.91, as verified across three independent random seeds (42, 7, 123) on the Kermany benchmark. The low standard deviation (accuracy std = 1.14%, AUC std < 0.004) confirms that results are statistically robust and reproducible.

A systematic binary-task baseline study demonstrates that DenseNet121 (91.03%, AUC = 0.9706) is the strongest standalone architecture, while EfficientNetB3 (73.88%) underperforms the custom CNN baseline—a practically useful negative finding for medical imaging model selection. Grad-CAM visualizations confirm that CBAM-DenseNet121 attends to anatomically plausible pulmonary regions for each class, supporting its potential for clinical deployment under appropriate radiologist supervision.

Limitations include: (1) the Kermany dataset originates from a single paediatric institution, limiting demographic generalisability; (2) bacterial/viral sub-labels are derived from filename metadata rather than independent clinical confirmation, introducing potential label noise; (3) statistical significance was assessed via three-seed repetition (seeds 42, 7, 123) rather than formal hypothesis testing. Future work will address these gaps by validating on CheXpert [3] and NIH ChestX-ray14 [1]; quantitative Grad-CAM evaluation against radiologist bounding-box annotations; model compression for edge deployment; and prospective clinical validation in Bangladeshi hospital settings.

## AUTHOR

UTSHO KUMAR DEY is currently pursuing the B.Sc. degree in Computer Science and Engineering from Northern University of Business and Technology Khulna, Bangladesh. His research interests include deep learning, medical image analysis, computer vision, and AI applications in healthcare. He has professional experience in Python programming instruction and web development, and holds certifications from NVIDIA (Generative AI with Diffusion Models, 2025), Google (Gemini Certified Educator, 2025), and IBM (Python for Data Science, 2025).